\documentstyle[aps,prl,epsf,multicol]{revtex}
\begin{document}

\title{Particle Dynamics in a Mass-Conserving Coalescence Process}

\author{Meesoon Ha$^{1}$, Hyunggyu Park$^{2}$, and Marcel den Nijs$^{1}$}
\address{$^1$ Department of Physics, University of Washington,
         Seattle, Washington 98195-1560, U.S.A.}
\address{$^2$ Department of Physics, Inha University,
         Inchon 402-751, Korea}
\date{\today}

\maketitle

\begin{abstract}

We consider a fully asymmetric one-dimensional diffusion model
with mass-conserving coalescence.
Particles of unit mass enter at one edge of the chain
and coalesce while performing a biased random walk towards the other edge
where they exit. The conserved particle mass acts as a passive scalar in
the reaction process
$A+A\rightarrow A$, and allows an exact mapping to a restricted
ballistic surface deposition model for which exact results exist.
In particular, the mass-mass correlation function is exactly known.
These results complement earlier exact results
for the $A+A\rightarrow A$ process without mass.
We introduce a comprehensive scaling theory for this process.
The exact analytical and numerical results confirm its validity.

\pacs{PACS numbers:}
\end{abstract}

\begin{multicols}{2}
\narrowtext
Diffusion-limited chemical processes are at the focus of recent research.
These are dynamic systems where the chemical reaction time scales are
short compared to those controlling spatial fluctuations in concentration.
The latter dominate the kinetics, in particular in low dimensions.
Such processes display dynamic scale invariance with scaling properties
that are robust, tend to be universal, and not sensitive to many details
of the actual dynamics at the microscopic level.
Simplified models are therefore able to catch the essence of the process.
Moreover, some of these models are accessible to exact solutions
in one dimension.

An example of these is the one-species coalescence process,
$A+A\rightarrow A$ \cite{A1,A2,A3,A4,A5}.
Exact results for this process were obtained recently using the
so-called inter-particle distribution function (IPDF) method,
also known as the method of empty intervals \cite{IPDF}.
This includes several versions of the model,
with and without external input of particles and in the presence or absence
of a diffusion bias along the chain \cite{MFA,AAA}.
In this letter we address the fully asymmetric diffusion case
where particles enter at one edge of the chain,
and coalesce when they meet each other,
while performing a driven random walk towards the other edge \cite{AAA}.
We enhance this model by assigning a mass to each particle, which is preserved
during each merging event.

Consider a linear chain with $L$ sites.
Particles of unit mass enter at the left boundary $x=1$,
diffuse to the right and coalesce when they meet, and ultimately
exit at the right boundary $x=L$. The diffusion of the particles along the
chain is totally biased. Choose a site $x$ at random.
If occupied, the particle at this site moves to the next site $x\to x+1$.
If site $x+1$ is already occupied, the two particles merge,
\begin{equation}
\label{dyn-rule}
m_x(t+1)=0 ~~{\rm and}~~ m_{x+1}(t+1)=m_{x+1}(t)+m_x(t)
\end{equation}
with $m$ the total mass of each particle. Total mass is conserved during
coalescence. Site $x=1$ is the input boundary. If chosen as an update site,
it is immediately refilled by the reservoir $m_1(t+1)=1$ and
$m_2(t+1)=m_2(t)+m_1(t)$ At the opposite edge, $x=L$, the particles simply
fall off the chain, $m_L(t+1)=0$, back into the reservoir.

The mass rides like a passive scalar on top of the particles.
It has no effect on the transition probabilities. We could describe the
process just as well in terms of only occupation numbers, $c_x=0,1,$
for empty or occupied site. The latter is the fully asymmetric
$A+A\rightarrow A$ process.

\begin{figure}
\centerline{\epsfxsize=6.5cm \epsfbox{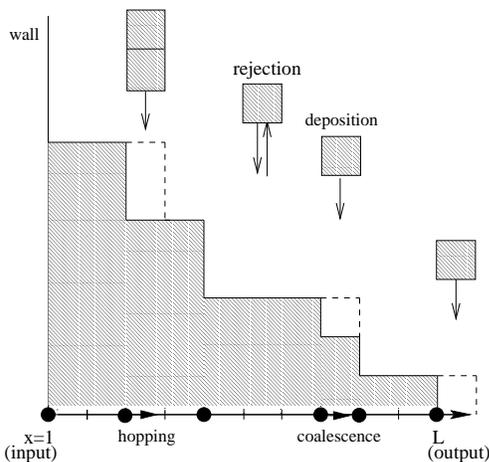}}
\caption
{Surface representation of the $A+A\rightarrow A$ process.
The step heights along the stairway represent the mass of the particles.
Each deposition process fills-up the entire step, such that the adjacent step
has a new height equal to the sum of the two.}
\label{stairway}
\end{figure}

The mass is a useful parameter. It allows an exact mapping onto
an exactly soluble restricted ballistic surface deposition (RBD)
model \cite{RBD}. Consider a one dimensional interface, as shown in
Fig.\ref{stairway}. The steps in the downward staircase can take any
magnitude $m_x=0,1,2,\cdots$. During each time step, one of the columns at
$x+\frac{1}{2}$ is chosen at random, $m_x$ particles are deposited onto it
such that the entire step fills up and the step at $x+1$ grows
to $m_{x+1}+m_x$. Exact results for this RBD model have been obtained
earlier using a generating function approach \cite{RBD}.
We can reinterpret those exact results in framework of the asymmetric
$A+A\rightarrow A$ reaction process.

Starting from the master equation, closed form recursive equations of
motion are obtained for the mass distribution along the chain
$M(x,t)= \langle m_x (t)\rangle$ and also (as discussed later) the
two-point mass correlations. $M(x,t)$ obeys the relation
\begin{equation}
\label{mass}
M(x,t+1) = (1-\frac{1}{L}) M(x,t) +\frac{1}{L}M(x-1,t)
\end{equation}
with boundary condition $M(1,t)=1$ (see eqs.(8)-(15) in reference
\cite{RBD} for details). This implies that in the stationary state the
mass distribution is uniform, $M(x)=1$, and there is a direct link between
particle concentration $C(x)=\langle c_x\rangle$ and
the average mass carried by each particle, $\tilde M(x)$.
They are exactly related as $\tilde M(x)= M(x)/C(x)$ where
$\tilde M(x)$ measures the average mass over occupied sites only.
Particle coalescence creates increasingly heavier particles
in the chain towards the right. At the same time they become more
sparsely spaced, because on average the mass remains constant along the
chain.

In the alternate RBD surface representation, this means that
although the step heights increase along the surface
from left to right, the steps occur at correspondingly larger intervals,
such that the stationary state average slope of the staircase remains
constant.

This dynamic process has a peculiar hierarchical structure.
Consider the path of one unit of mass. It performs a biased random walk,
but completely uncorrelated from all other masses. Similarly,
two specific mass units perform completely independent random walks,
until the moment they meet. From there on they move randomly but as a
bound pair. Again, all other particles play no role.

This hierarchical property suggests the following scaling theory.
Consider a specific unit of mass. It moves to the right with average
velocity $v=1$. The standard deviation in its position is proportional to
the square root of the time and distance traveled, i.e. $\Delta r\sim
x^{\frac{1}{2}}$. While diffusing to the right this particle merges with
other masses. The total amount of mass it sweeps up is expected then to be
of order $\Delta r$, i.e., that the average mass of occupied sites scales
as $\tilde M(x)\sim x^{\frac{1}{2}}$. The amount of swept-up mass does
not depend on the particle concentration $C(x)$, because
although the merging events reduce in frequent, they increase in size.
However, this increasing lumpiness will show in increasing
statistical fluctuations along the chain in Monte Carlo simulations.

Next, we predict that the above random walk exponents are exact. This
presumes the absence of intricate correlations between the particles.
The hierarchical structure of the equations of motion, and the diffusion
equation structure of recursion relations in both exact solution methods
suggest this prediction. In the following we demonstrate the validity of
this scaling theory from both numerical and analytical exact results.

Hinrichsen {et al.}\cite{AAA} obtained the exact stationary state
particle concentration. It scales as
\begin{equation}
\label{av-density}
C(x) =
\sqrt{\frac{1}{\pi x}}[1+\frac{3}{8x}]+O(x^{-5/2}).
\end{equation}
This agrees with our scaling theory, because of the exact relation
$\tilde M(x)= M(x)/C(x)$. The scaling argument predicts that $\tilde M(x)$
grows as $x^{\frac{1}{2}}$, while eq.(\ref{mass}) implies that
$M(x)=1$.

\begin{figure}
\centerline{\epsfxsize=6.5cm \epsfbox{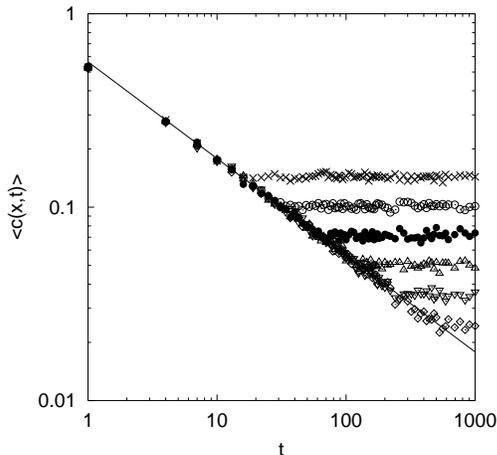}}
\caption
{The time dependent particle concentration $\langle c(x)\rangle$
as function of time at various values of the distance from the source,
$x=16(\times), 32(\circ), 64(\bullet),
128(\bigtriangleup), 256(\bigtriangledown)$, and
$512(\diamond)$ at chain length
$L=512$. The slope of the drawn line is -1/2.}
\label{part}
\end{figure}

Figure \ref{part} illustrates the above numerically. It shows the time
evolution of the particle density at various values of $x$ starting from
the uniform initial state where every site is occupied by one unit
of mass particle. We averaged over 10 independent Monte Carlo runs.
Initially the curves coincide, until the time when particles
from the input edge reach that specific site. As expected
this crossover time scales linearly with $x$, i.e., with the uniform
average particle velocity along the chain. The slope of the initial curve
is $C(x,t)\sim t^{-\frac{1}{2}}$, and the stationary state plateau values
scale as $C(x) \sim x^{-\frac{1}{2}}$, both in accordance with the scaling
theory. Notice also that the statistical fluctuations increase
with the distance from the source $x$.

\begin{figure}
\centerline{\epsfxsize=6.5cm \epsfbox{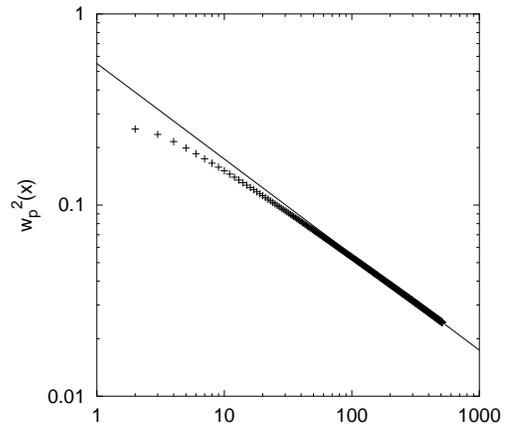}}
\caption
{fluctuations in the particle concentration from
the auto-correlation functions $w_{p}^{2}(x)=\langle c^{2}(x)
\rangle-\langle c(x)\rangle^{2}$ at chain length $L=512$. The slope of
the drawn line is -1/2.}
\label{W-p}
\end{figure}

The mass auto-correlation function,
$w_m^2=\langle m_x^2\rangle - \langle m_x\rangle^2$ measures these
fluctuations, and is a special case of the two-point correlator
discussed below for which we have an exact solution from
the mapping to the RBD model. At large $x$ the fluctuations grow as
$w_m(x)\simeq \sqrt x$, in accordance with our random-walk based scaling
theory. We checked numerically the fluctuations in the particle density.
Fig.\ref{W-p} shows that they decay as $w_p \simeq A/\sqrt x$,
again consistent with the scaling theory.

Consider the particle-particle correlation function, $g_p (x,r)=
\langle c_xc_{x+r}\rangle -\langle c_x\rangle \langle c_{x+r}\rangle$
and the mass-mass correlation function, $g_m (x,r)=
\langle m_xm_{x+r}\rangle-\langle m_x\rangle \langle m_{x+r}\rangle$.
As far as we know, no exact results are available for $g_p (x,r)$,
from the method of empty intervals, although it seems within reach.
On the other hand, the two-point mass correlator
$G_m(x,y,t)= \langle m_x (t) m_y (t) \rangle$
was shown in \cite{RBD} to obey in the stationary steady state the
recursion relations
\begin{eqnarray}
& G_m(1,x)   & =G_m(1,x-1),\nonumber\\
& G_m(x,x)   & =2G_m(x-1,x)+G_m(x-1,x-1),\nonumber\\
& G_m(x-1,x) & =\frac{1}{2}G_m(x-2,x),\nonumber\\
& G_m(x,y)   & =\frac{1}{2}[G_m(x-1,y)+G_m(x,y-1)],
\end{eqnarray}
where $|x-y|\ge 2$.
This set of coupled equations can be solved exactly,
\begin{eqnarray}
\label{stst-mass-corr}
& G_m(1,x)   & =1,\nonumber\\
& G_m(x,x)   & =1+4(x-1)Z_{2(x-1)}(0),\nonumber\\
& G_m(x-1,x) & =Z_{2(x-2)}(0),\nonumber\\
& G_m(x,y)   & =\sum_{n=0}^{r-1}Z_{2(y-2)}(n) =1-\sum_{n=r}^{y-2}Z_{2(y-2)}(n),
\end{eqnarray}
with $r\equiv (y-x) >0$, and
\begin{equation}
Z_{2p}(n)=\frac{(2p-n)!}{p!(p-n)!}(\frac{1}{2})^{2p-n},
\end{equation}
which is related to the probability that a random walker returns to
its starting point exactly $n$ times up to $2p$ steps, see \cite{RBD} for
details.

\begin{figure}
\centerline{\epsfxsize=6.5cm \epsfbox{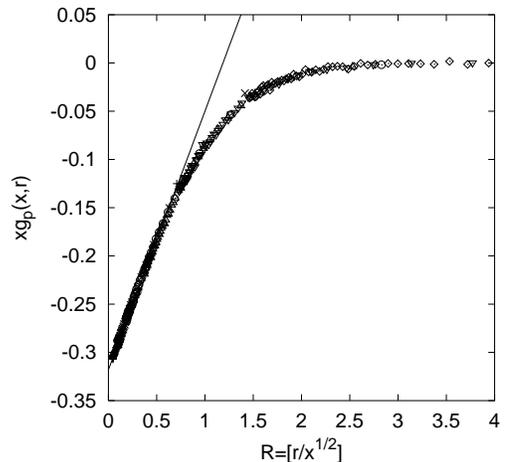}}
\caption
{The scaling function for the particle-particle correlator $g_p$,
from 10 independent Monte Carlo runs each averaged over $10^6$ time
steps at chain length $L=512$ for various distances from the source,
$r=1(+), 2(\times), 4(\circ), 8(\bigtriangleup),
16(\bigtriangledown)$, and $32(\diamond)$. 
The drawn line is obtained from eq.(\ref{fp}).}
\label{g-p}
\end{figure}

\begin{figure}
\centerline{\epsfxsize=6.5cm \epsfbox{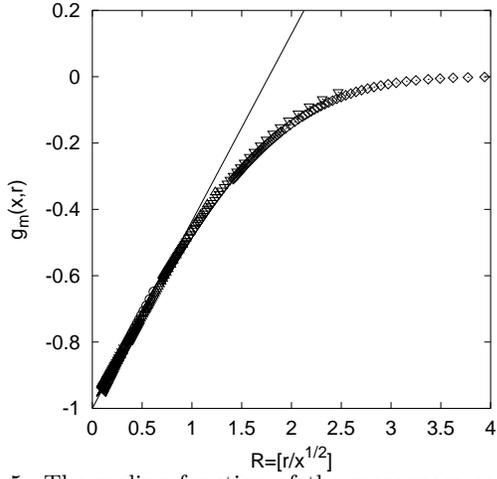}}
\caption
{The scaling function of the mass-mass correlator $g_m$, from
enumeration of the exact recurrence relations, at chain length $L=512$
for various distances from the source, $r=2(\times), 4(\circ),
8(\bigtriangleup), 16(\bigtriangledown)$, and $32(\diamond)$. 
The drawn line is the tangent at small $R$, see between 
eq.(\ref{corr}) and eq.(\ref{fp}).}
\label{g-m}
\end{figure}

We expect the following scaling form for both  $g_m$ and $g_p$
in the stationary state:
\begin{equation}
\label{g-scaling}
g_\alpha (x,r)=b^{-2{\rm x}_\alpha} g_\alpha (b^{-1}x, b^{-1/2}r),
\end{equation}
with $b$ an arbitrary scale factor. The distance $r$ from the source plays
the role of time in our diffusion-coalescence type scaling argument.
Therefore $r$ should scale with respect to the correlator distance $x$
as $r\sim \sqrt{x}$. The other exponents, ${\rm x}_p$ and ${\rm x}_m$,
follow by power-counting. They must be the dimensions of the particle and
mass
concentrations, $C(x)$ and $M(x)$, which are equal to
${\rm x}_p =\frac{1}{2}$
and ${\rm x}_m = 0$ according to the previous discussion.
We perform numerical simulations for $g_p$, and exact enumerations of the
exact formula for $g_m$. Fig.\ref{g-p} and Fig.\ref{g-m}
show the scaling functions $f_p$ and $f_m$, defined as
\begin{eqnarray}
\label{corr}
g_p=x^{-1} f_p (r/\sqrt{x}),\nonumber\\
g_m=f_m(r/\sqrt{x}),
\end{eqnarray}
The data collapses perfectly and thus confirms the validity of
the scaling relations, eq.(\ref{g-scaling}).

Both scaling functions vanish at large $R=r/\sqrt{x}$, as they should.
At small $R$ they are both linear in the scaling variable.
It is easy to evaluate the exact recursion relations
for $G_m$ in the limit of large $x$ and fixed small $r$.
This yields $f_m(R)\simeq -1+ \frac{1}{\sqrt{\pi}} R$.

The particle correlator scaling function is also linear at small $R$.
We find numerically that
\begin{equation}
\label{fp}
f_p(R)\simeq -a+ b R
\end{equation}
with $a=0.318(1)$ and $b=0.268(6)$ and we suspect that the overall
factor $a=1/\pi$.

The above numerical and exact results are all in full agreement with the
simple scaling picture where we treat the particles as performing
free independent biased random walks before they merge. None of
the scaling exponents differ from their diffusion values.
The final verification for the validity of this intuitive explanation
is the shape of the stationary state mass distribution function,
$p(m,x)$, i.e.,  the probability to find a particle of mass $m$
at a site $x\gg 1$.
Our scaling theory presumes that this distribution behaves in the
same manner as the probability for a specific tagged particle to reach
site $x$ with mass $m$. Each tagged particle follows an independent biased
random walk and its mass grows proportional to the spatial fluctuations
about its average path.

\begin{figure}
\centerline{\epsfxsize=6.5cm \epsfbox{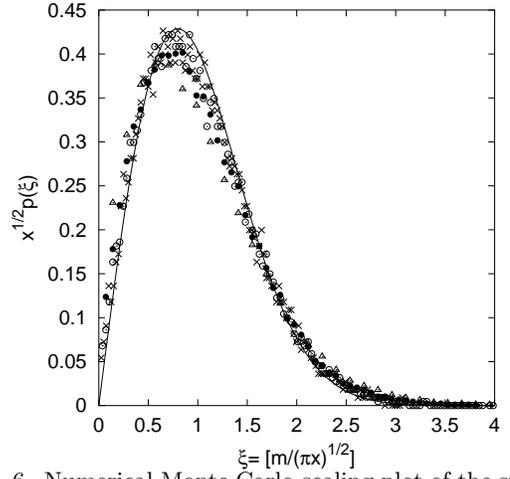}}
\caption
{Numerical Monte Carlo scaling plot of the stationary state
distribution $p(m,x)$, the probability to find a particle of mass $m$ at
site $x$, for various $x=16(\times), 64(\circ),
256(\bullet)$, and $512(\triangle)$. We plot
$\sqrt{x}p(x,m)$ versus
$\xi=m/\sqrt{\pi x}$, the Gaussian form of eq.(\ref{m-prob}).}
\label{dist}
\end{figure}

Fig.\ref{dist} shows  numerical results for $p(m,x)$
at system size $L$=512 for various values of $x$=16, 64, 256, and 512.
It is obtained from 10 independent runs, each  $10^6$ Monte Carlo steps long.
The data is plotted in terms of $\sqrt{x} p(m,x)$ versus the scaling variable
$\xi=m/\sqrt{\pi x}$ and collapses well onto Gaussian with a linear
prefactor,
\begin{equation}
\label{m-prob}
p(m,x)= A ~\frac{m}{x} ~\exp (-Bm^2/x).
\end{equation}
The parameters $A$ and $B$ are
predetermined by the normalization condition and the
scaling of the particle density (see eq.(\ref{av-density}))
\begin{eqnarray}
\sum_{m=0}^{\infty}p(m,x)=1,\nonumber\\
\tilde M(x) =\sum_{m=0}^{\infty}mp(m,x)=1/C(x),
\end{eqnarray}
which yields $A=\frac{1}{2}$ and $B=\frac{1}{4}$.
The drawn curve in  Fig.\ref{dist} corresponds to this Gaussian with the
above values of $A$ and $B$.
Eq.(\ref{m-prob}) is the simplest Gaussian form consistent with the
requirement that $p(m,x)$ vanishes in the limit $m\to 0$.
According to our intuitive picture,
$p(m,x)$ is related to the probability that a  biased random walker
with drift velocity $v_d=\frac{1}{2}$,
makes an excursion of size $m$ from its average path
before reaching site $x$,  irrespective of (i.e., averaged over all)
the starting times at site $x=1$.

In conclusion, we present new exact results for the
$A+A\rightarrow A$ type coalescense process by introducing mass to the
particles, and exploring the exact mapping to a surface deposition model,
the so-called RBD type surface growth model. In addition, we propose a
scaling theory based on the assumption we can treat the particles as
performing
free independent biased random walks before they merge. All scaling
exponents should
then take naive diffusion values. The above numerical and exact results are all
in full agreement with this random walk type scaling. It appears therefore
that the scaling properties of $A+A \rightarrow A$ type dynamics are now fully
understood, and actually, in the final analysis, are predictable
from random walk considerations only.

\section*{Acknowledgments}
This work is supported by NSF grant DMR-9700430 and by the Korea Research
Foundation (98-015-D00090).

\newpage

\end{multicols}
\end{document}